\documentclass[journal=NanoLetters,manuscript=article]{achemso}
\setkeys{acs}{keywords = true}
\usepackage{chemformula} % Formula subscripts using \ch{}
\usepackage[T1]{fontenc} % Use modern font encodings
\RequirePackage{lineno}
\usepackage{xcolor}
\usepackage{caption}
\usepackage{multirow}
\usepackage{color}
\usepackage{bm}
\usepackage{braket}
\usepackage{amsmath, amssymb, amsfonts}
\usepackage{calc}

\author{Tianyun Lin$^{\infty}$}
\affiliation
{State Key Laboratory of Low Dimensional Quantum Physics and Department of Physics, Tsinghua University, Beijing 100084, P. R. China}
\author{Yongkang Ju$^{\infty}$}
\affiliation
{School of Materials Science and Engineering, Beihang University, Beijing 100191, P. R. China}
\author{Haoyuan Zhong}
\affiliation
{State Key Laboratory of Low Dimensional Quantum Physics and Department of Physics, Tsinghua University, Beijing 100084, P. R. China}
\author{Xiangyu Zeng}
\affiliation
{Department of Physics and Beijing Key Laboratory of Opto-electronic Functional Materials and Micro-nano Devices, Renmin University of China, Beijing 100872, P. R. China}
\author{Xue Dong}
\affiliation
{Department of Physics and Beijing Key Laboratory of Opto-electronic Functional Materials and Micro-nano Devices, Renmin University of China, Beijing 100872, P. R. China}
\author{Changhua Bao}
\affiliation
{State Key Laboratory of Low Dimensional Quantum Physics and Department of Physics, Tsinghua University, Beijing 100084, P. R. China}
\author{Hongyun Zhang}
\affiliation
{State Key Laboratory of Low Dimensional Quantum Physics and Department of Physics, Tsinghua University, Beijing 100084, P. R. China}
\author{Tian-Long Xia}
\affiliation
{Department of Physics and Beijing Key Laboratory of Opto-electronic Functional Materials and Micro-nano Devices, Renmin University of China, Beijing 100872, P. R. China}
\author{Peizhe Tang}
\affiliation
{School of Materials Science and Engineering, Beihang University, Beijing 100191, P. R. China}
\alsoaffiliation
{Max Planck Institute for the Structure and Dynamics of Matter, Center for Free-Electron Laser Science, 22761 Hamburg, Germany}
\email{peizhet@buaa.edu.cn}
\author{Shuyun Zhou}
\affiliation
{State Key Laboratory of Low Dimensional Quantum Physics and Department of Physics, Tsinghua University, Beijing 100084, P. R. China}
\alsoaffiliation
{Collaborative Innovation Center of Quantum Matter, Beijing 100084, P. R. China}
\email{syzhou@mail.tsinghua.edu.cn}

\title
{Ultrafast Carrier Relaxation Dynamics in a Nodal-Line Semimetal PtSn$_4$}


\begin{document}



\begin{abstract}
Topological Dirac nodal-line semimetals host topologically nontrivial electronic structure with nodal-line crossings around the Fermi level, which could affect the photocarrier dynamics and lead to novel relaxation mechanisms. Herein, by using time- and angle-resolved photoemission spectroscopy, we reveal the previously-inaccessible linear dispersions of the bulk conduction bands above the Fermi level in a Dirac nodal-line semimetal PtSn$_4$, as well as the momentum and temporal evolution of the gapless nodal lines. A surprisingly ultrafast relaxation dynamics within a few hundred femtoseconds is revealed for photoexcited carriers in the nodal line. Theoretical calculations suggest that such ultrafast carrier relaxation is attributed to the multichannel scatterings among the complex metallic bands of PtSn$_4$ via electron-phonon coupling. In addition, a unique dynamic relaxation mechanism contributed by the highly anisotropic Dirac nodal-line electronic structure is also identified. Our work provides a comprehensive understanding of the ultrafast carrier dynamics in a Dirac nodal-line semimetal.
\end{abstract}


Three-dimensional (3D) topological semimetals\cite{Armitagereview2018RMP,dinghongreviewRMP}, such as Dirac, Weyl or nodal-line semimetals, have attracted extensive attention due to their exotic electronic structures containing gapless crossing points and nontrivial topological properties. Such topological semimetals are characterized by the crossing of the conduction band and the valence band near the Fermi energy $E_F$ at isolated points (Dirac or Weyl points) or along a 1D curve (nodal line) in the 3D Brillouin zone (BZ). The crossing points or lines cannot be gapped by any perturbation that preserves a certain symmetry group of the target material, and therefore, they are topologically protected by the symmetry group and can be associated with a topological invariant\cite{Armitagereview2018RMP,FangzhongreviewCPB_2016}. The exotic electronic and topological properties lead to novel optical response\cite{weberreview2021jap,sundong2020semimetals,ma2021topology,BCHreview2022NRP,Dongbin2024}, for example, circular photogalvanic effect and shift current in the Weyl semimetal\cite{TaAsCPGE2017np, TaAsshiftcurrent}, enhanced second-harmonic generation signal in the Weyl semimetal\cite{JosephSHG2017np}, and high harmonic generation in the Dirac semimetal\cite{CdAsHHG2020PRL}. Furthermore, the conical electronic structure of topological semimetals may lead to distinctive carrier relaxation dynamics. For example, light-induced population inversion has been reported in 3D Dirac semimetal Cd$_3$As$_2$ \cite{bchpopulationinversion2022nl}, and different relaxation processes of bulk and surface states have been reported in a nodal-line semimetal ZrSiS \cite{NeupaneZrSiS2022cp,MazharZrSiS2018APL}. Thus, revealing the complete topological electronic structure including the conduction bands above $E_F$ as well as the band-resolved carrier relaxation dynamics upon photoexcitation is important for a comprehensive understanding of topological semimetals, and for advancing their applications in electronics and optoelectronics.

PtSn$_4$ is a nodal-line semimetal, which exhibits novel transport properties\cite{firsttransport2012PRB, PHEYan2020JPCM, HarimaPS2018PBCM, ZhangYHPSdhva2018JPCM, SunYPPSLifshtz2018PRB, Eisterer2019JOEATP,fuchenguang2020Research} such as extremely large magnetoresistance\cite{firsttransport2012PRB}, and planar Hall effect\cite{PHEYan2020JPCM}. Two types of nodal lines have been discovered in previous works\cite{kaminskiARPES2016np}: one is below $E_F$ (-60 meV) around the X point which has been well identified; another nodal line around the Z point crosses $E_F$, some part of which is even above $E_F$. The latter nodal line may play a critical role in the transport properties as well as optoelectronic applications because it is close to the Fermi level. However, so far, experimental information of the nodal-line electronic structure around the Z point is rather limited, because angle-resolved photoemission spectroscopy (ARPES) can probe only the occupied states below $E_F$. How the Dirac nodal-line electronic structure above $E_F$ evolves in the momentum space, and how such nodal-line electronic structure affects the carrier relaxation dynamics are important questions to address.

Here, by photoexciting electrons into the conduction bands above $E_F$ and subsequently probing the transient electronic structures using time-resolved ARPES (TrARPES) measurements, we reveal the dispersions of the conduction bands above $E_F$ and the momentum evolution of the gapless nodal lines projected onto the side surface, thereby clarifying the bulk nodal-line electronic structure around the Z point. Moreover, band-resolved relaxation dynamics of the photoexcited carriers are revealed, which show a surprisingly fast relaxation process with a lifetime as short as a few hundred femtoseconds. Combining experimental results with calculated momentum-resolved electron-phonon (el-ph) coupling matrix using the density functional theory (DFT) calculations, the fast relaxation is attributed to interband scatterings between these metallic bands in the complex Fermi surface of PtSn$_4$. Furthermore, we identify a dynamical relaxation mechanism which notably is absent in Dirac and Weyl semimetals yet distinctive in the Dirac nodal-line semimetal, wherein the pivotal role is played by acoustic phonons. 
Our work not only provides a more complete picture of the nodal-line electronic structure of PtSn$_4$ but also identifies the roles of the Dirac nodal-line electronic structure in the ultrafast carrier relaxation dynamics.

\begin{figure*}[htbp]     
	\centering                                                            
	\includegraphics{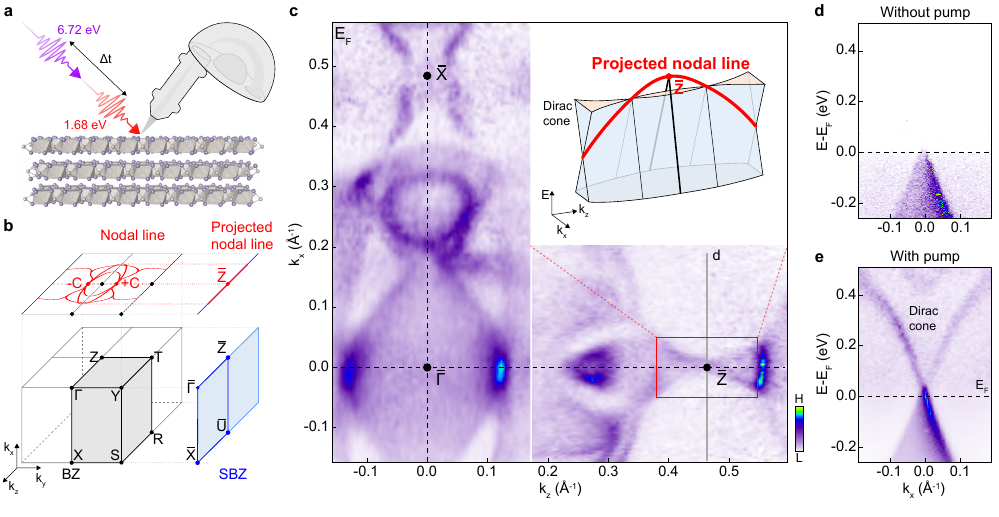}                                          
	\caption{Schematic of the experimental setup and overview of the electronic structure of PtSn$_4$. (a) Schematic of the TrARPES experimental setup and the crystal structure of PtSn$_4$. The pump photon energy is 1.68 eV and the probe photon energy is 6.72 eV. (b) Brillouin zone (BZ), and the projected surface Brillouin zone (SBZ) of PtSn$_4$. The upper panel is the schematic for the nodal line in the bulk BZ and the projected nodal line in SBZ. (c) Fermi surface map of PtSn$_4$. The inset shows a schematic illustration of the projected Dirac nodal-line structure near the $\overline{\rm Z}$ point. (d) ARPES dispersion image measured by cutting through the $\overline{\rm Z}$ point (indicated by the gray line in (c)). (e) TrARPES dispersion image measured along the same direction, at delay time $\Delta t$ = 0.1 ps. In order to clearly visualize the dispersion above $E_F$, the intensity at each energy is normalized by the integrated intensity over the full momentum range.}                                                                                                 
	\label{fig1}   
\end{figure*}
                                                                                                            
Figure~1a shows a schematic illustration of the TrARPES setup and the crystal structure of PtSn$_4$. The BZ of PtSn$_4$ is shown in Figure~1b, where calculated nodal lines close to the $\rm Z$ point and their corresponding projections onto the surface BZ (SBZ) are also plotted. 
Because of the C$_2$ rotational symmetry and mirror symmetry, the 3D bulk electronic states are 4-fold degenerate along the Z-T direction. In particular, band structure calculation shows two 4-fold crossing points at the $\pm C$ points, and four gapless nodal lines at different energies on the plane of ZT$\Gamma$Y in the 3D BZ (red curves in the upper panel of Figure~1b, see more details in Figure~S1). When projecting onto the (010) surface BZ (SBZ, marked by the blue color in Figure~1b) which is measured by ARPES experiments, these nodal lines correspond to a line near the $\overline{\rm Z}$ point (red line in the upper-right corner of Figure~1b), where the $\pm C$ points are projected to the $\overline{\rm Z}$ point exactly. Under such projection, bulk gapless points with the same $k_y$ around the $\pm C$ points but belonging to different nodal lines will project onto the same point in SBZ, resulting in the Dirac crossings with a small energy variation (see the schematic illustration in the inset of Figure~1c). Direct experimental observation of the projected nodal-line electronic structures near the $\overline{\rm Z}$ point is important for verifying the nodal-line electronic structure and revealing the ultrafast dynamics, which are the main focus of this work.

Figure~1c-e shows an overview of the experimental electronic structure of PtSn$_4$, with a special focus on the electronic structure around the $\overline{\rm Z}$ point. The experimentally measured Fermi surface map in Figure~1c shows many complex pockets, consistent with previous static ARPES measurements\cite{kaminskiARPES2016np}. Cutting through the $\overline{\rm Z}$ point, a Dirac cone-like valence band is identified in Figure~1d, where the crossing point at the $\overline{\rm Z}$ point is slightly above $E_F$ and thus beyond ARPES detection. Upon laser pumping, the conduction band above $E_F$ is transiently populated by photoexcited electrons, making the conical dispersion with crossing point above $E_F$ observable experimentally (Figure~1e). Therefore, by performing TrARPES measurements, we can obtain a more complete picture of the Dirac cone-like band, revealing not only the valence band dispersion but also the conduction band dispersion above $E_F$.

\begin{figure*}[htbp]
	\centering
	\includegraphics{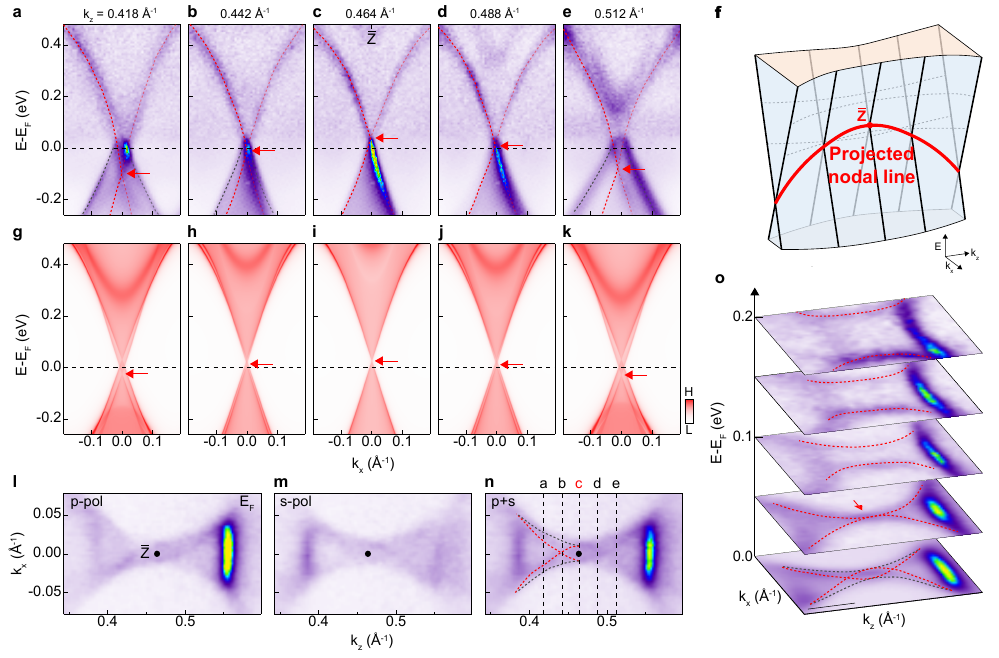}
	\caption{Fine electronic structure of the projected Dirac nodal line near the $\overline{\rm Z}$ point. (a-e) Dispersion images measured along dashed lines in (n) at delay time $\Delta t$ = 0.1 ps. Red and gray dashed curves are guides for the dispersion of the two sets of cone-like bands. In order to clearly visualize the dispersion above $E_F$, the intensity at each energy is normalized by the integrated intensity over the full momentum range. (f) Schematic summary of the projected Dirac nodal line near the $\overline{\rm Z}$ point. (g-k) Calculated spectra corresponding to data shown in (a-e). The spectral weight is plotted in the logarithmic scale for a clear visualization. (l, m) Fermi surface maps measured with $p$-polarized probe beam (l) and $s$-polarized probe beam (m). (n) Summing up of data in (l) and (m) to show clearly the twisted electronic structure at the $\overline{\rm Z}$ point. Red and gray dashed curves are guides for the line nodes on the Fermi surface from two sets of cone-like bands. (o) Three-dimensional electronic structure at the $\overline{\rm Z}$ point above $E_F$ at delay time $\Delta t$ = 0.1 ps. The red and gray dashed curves indicate the energy contours, and the arrow points to the crossing point. The scale bar represents 0.1 \AA$^{-1}$.}
	\label{fig2}
\end{figure*}

To further reveal how the crossing point evolves in momentum space, Figure~2 shows dispersion images measured near the $\overline{\rm Z}$ point. Figure~2a-e shows the evolution of the full set of electronic structures along momentum cuts perpendicular to the projected nodal line (indicated by black dashed lines in Figure~2n) upon laser pumping (see the raw data in Figure~S2).
Away from the $\overline{\rm Z}$ point, two sets of cone-like bands are observed in Figure~2a,~b,~e (red and gray dashed curves). While at the $\overline{\rm Z}$ point, these two sets of cones merge into one in Figure~2c, in agreement with the above symmetry analysis and DFT calculations, confirming that the conical bands are only degenerate at the $\overline{\rm Z}$ point.
Below we mainly focus our discussions on the cone-like band indicated by red dashed curves, which is observed clearly even when moving away from the $\overline{\rm Z}$ point. The Dirac crossing point gradually moves down from above the Fermi level (Figure~2c) to below the Fermi level (Figure~2a,~b,~e) as pointed out by the red arrows, confirming that these projected bulk curved nodal lines are gapless and they cross the Fermi level, as schematically illustrated in Figure~2f. DFT calculated results shown in Figure~2g-k reveal similar surface band structures, in which bulk gapless points from different Dirac nodal lines overlap with each other and form the Dirac cone-like crossings (marked by the red arrows) when projected onto the side surface.

The formation of curved projected nodal lines is also supported by the experimental 3D electronic structure shown in Figure~2l,~m. In particular, the application of different probe polarizations allows us to resolve the multiple pockets around the $\overline{\rm Z}$ point with different polarization contrast. By summing up these two maps, the full electronic structure can be obtained in Figure~2n, where band crossings from the projected nodal lines are identified (indicated by the red and gray dashed curves). When moving to higher energy in Figure~2o, the separated band crossing points merge to one above the Fermi level (indicated by the red arrow) and finally disappear at even higher energy, again supporting the curved projected nodal line schematically illustrated in Figure~2f.

\begin{figure*}[htbp]
	\centering
	\includegraphics{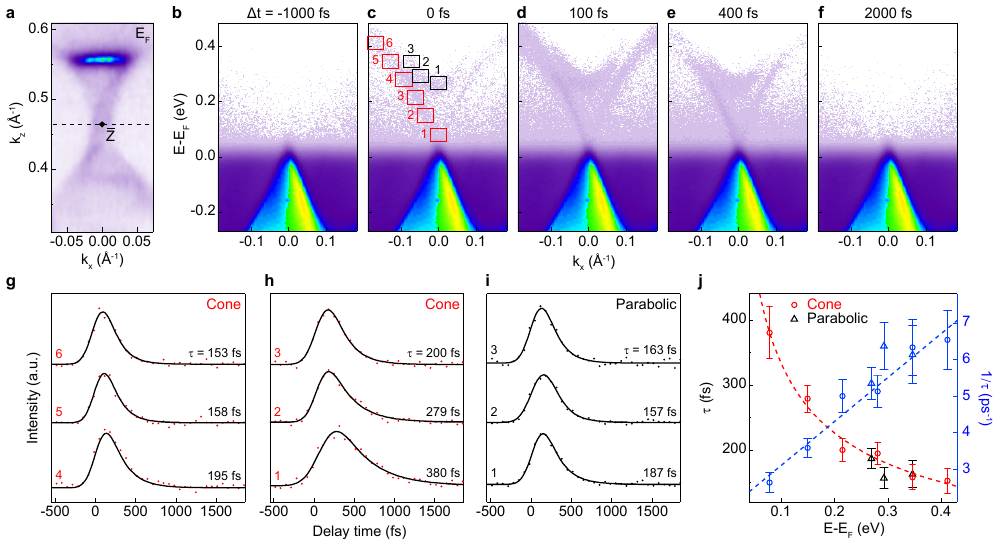}
	\caption{Ultrafast relaxation dynamics of photoexcited carriers. (a) Fermi surface map at the $\overline{\rm Z}$ point, where the dashed line indicates the momentum direction along which the dispersion images are measured. (b-f) Dispersion images measured across the $\overline{\rm Z}$ point at different pump-probe delay times. The dispersion images are acquired at a fixed angle to make sure it cuts the $\overline{\rm Z}$ point at Fermi energy. (g-i) Temporal evolution of the photoexcited electron intensity at different energies, obtained by integrating the spectrum intensity over boxes indicated in (c) for cone-like band (g,~h) and parabolic band (i). (j) Extracted relaxation times for the cone-like band (red circles) and parabolic band (black triangles), and corresponding scattering rates (blue circles and blue triangles) of the photoexcited electrons at different energies.}
	\label{fig3}
\end{figure*}

TrARPES measurements not only allow revealling of electronic structures of the unoccupied states but also allow capturing of the ultrafast carrier relaxation dynamics, which is critical for understanding the microscopic scattering events and the corresponding many-body interactions.
Figure~3b-f shows TrARPES snapshots at different delay times of the cone-like band measured by cutting through the $\overline{\rm Z}$ point, as shown in Figure~3a. Upon laser pumping, electrons are excited to unoccupied bands and quickly thermalized as hot electrons. At $\Delta$t = 0, populated electrons are observed in the upper branch of the cone-like band above the Fermi level (marked by red boxes in Figure~3c) and a near parabolic conduction band at higher energy (marked by black boxes in Figure~3c), both of which relax to the equilibrium state within 2 ps. The continuous temporal evolution of photoexcited electrons at different energy and momentum positions is also shown in Figure~3g-i, further revealing the details of the band-resolved photoexcitation and relaxation processes. The corresponding relaxation time $\tau$ is extracted by fitting the time traces using the product of a step function and an exponential decay function convoluted with a Gaussian function\cite{bch2021timeresolutionRSI, ZHY100FSRSI2022} (see details in Methods). The extracted relaxation time for the cone-like band at different energies is plotted in Figure~3j, which shows that the relaxation time ranges from 153 to 380 fs. For the parabolic conduction band at higher energy, the relaxation time is shorter, ranging from 163 to 187 fs. We note that the longest relaxation time of 380 fs for the cone-like band near the Fermi level is still much shorter compared to the few ps relaxation times reported in ZrSiS\cite{NeupaneZrSiS2022cp} and many other topological materials (see Table 1), such as Bi$_2$Se$_3$\cite{shenZXBS2012PRL,GedikBS2012PRL}, Sb$_2$Te$_3$\cite{HoferSbTe2014PRB,KimuraSbTe2015scirep}, MnBi$_{2n}$Te$_{3n+1}$\cite{zhyMBT2021nl,majchrzakMBT2023nl}, and Cd$_3$As$_2$\cite{bchpopulationinversion2022nl,gongqihuangCdAs2022JPCC}. The relaxation dynamics reported here are surprisingly fast, meanwhile, the scattering rate shows an overall linear dependence on the energy (Figure~3j), which is similar to the case of Dirac semimetal Cd$_3$As$_2$\cite{bchpopulationinversion2022nl}. We note that the reported relaxation time of graphene is also short (about 800 fs)\cite{Andreagraphene2013nm,Andreagraphene2015}, which results from the strong el-ph coupling and an additional Auger recombination process evolved in the relaxation process\cite{Polinigraphene2013nc,Andreagraphene2015}. Whether the fast relaxation process in PtSn$_4$ has the same origin as that in graphene needs further exploration.

\begin{table*}
	\caption{Dynamics of Different Topological Materials}
	\label{tbl1}
	\begin{tabular}{c|c|c|c|c|c}
		\hline
		Materials & PtSn$_4$ (this work) & Bi$_2$Se$_3$\cite{shenZXBS2012PRL,GedikBS2012PRL} & Sb$_2$Te$_3$\cite{HoferSbTe2014PRB,KimuraSbTe2015scirep} & MnBi$_{2n}$Te$_{3n+1}$\cite{zhyMBT2021nl,majchrzakMBT2023nl} & 	Cd$_3$As$_2$\cite{bchpopulationinversion2022nl,gongqihuangCdAs2022JPCC} \\ 
		\hline				
		$\tau$ (ps) & 0.4 & 6 & 1-2 &0.5-2 &6-7 \\
		\hline
	\end{tabular}
\end{table*}

To understand why the carrier relaxation is so fast in PtSn$_4$, we have performed a detailed theoretical calculation and analysis on electronic structures around the nodal lines and related scattering events for PtSn$_4$. The relaxation of hot electrons is dominated by the el-ph coupling, in which the coupling strength is strongly affected by electronic structures around the Fermi level (Figure~4b).
To evaluate the el-ph coupling strength, we choose one state on the Dirac nodal lines (P point shown in Figure~4a,b) as the initial state to calculate momentum-resolved el-ph coupling matrix elements $\lambda_{q \nu P}$ in primitive BZ (Figure~4a and Figure~S3), where $q$ and $\nu$ stand for the wave vector and the branch of a phonon mode, respectively.  The final state of the el-ph scattering matrix is in the energy window marked by the blue shade in Figure~4b, whose value is 2 times of largest phonon energy for PtSn$_4$. As calculated results show in Figure~4c, in addition to the large contributions from optical phonon modes near the $\Gamma$ point, some other phonons with large $q$ values also contribute significantly to el-ph scattering. This suggests that metallic bands away from the nodal lines could also act as final states for hot electron relaxations and contribute to an ultrafast relaxation process with a relaxation times cale of $\tau \sim 10^2$ fs, as schematically illustrated by red curved arrows in Figure~4b.

\begin{figure*}[htbp]
	\centering
	\includegraphics{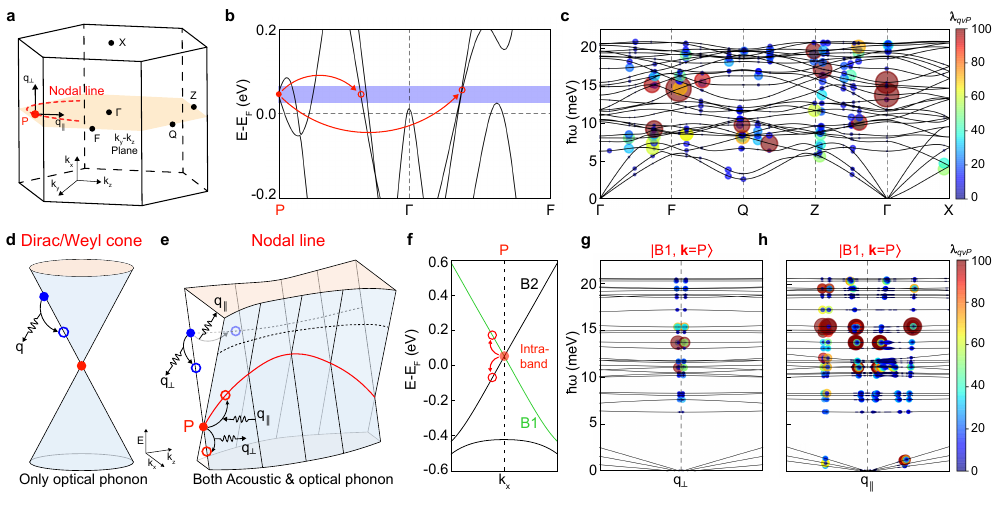}
	\caption{Electron-phonon coupling matrix for states around the Dirac nodal lines mapping onto phonon bands of PtSn$_4$. (a) First BZ of the primitive cell of PtSn$_4$ and schematic illustration of the nodal line. (b) Electronic structure near $E_F$. Constrained by the energy conservation, el-ph scattering only occurs in the blue-shade energy window. (c) Phonon dispersions for PtSn$_4$ and the momentum-resolved el-ph coupling matrix. The initial state is a crossing point in the nodal line (labeled as P), and the final states are within the blue-shaded energy window in (b). The size and color of the circles both represent the el-ph coupling strength $\lambda_{q \nu k}$. (d,~e) Schematic illustration for el-ph scattering in the 3D Dirac/Weyl cone (d) and the Dirac nodal line (e). (f) Cone-like band structures along $k_x$ with two bands (labeled as B1 and B2) crossing each other. (g,~h) Phonon dispersions for PtSn$_4$ near the $\Gamma$ point. The el-ph coupling matrix $\lambda_{q\nu P}^{intra}$ for intraband scattering is mapped on phonon bands along directions of perpendicular (g) and parallel (h) to the nodal line, which is scattered by different phonon modes initially from B1 state at P point.}
	\label{fig4}
\end{figure*}

	In addition to the el-ph coupling among the complex metallic bands, the unique nodal-line electronic structure also contributes to the ultrafast relaxation process of PtSn$_4$. Considering the 1D crossing curve in the Dirac nodal-line semimetal, we argue the relaxation mechanism scattered by el-ph coupling in the nodal line has a fundamental difference from those in Dirac and Weyl semimetals. For states around the Dirac and Weyl points with linear dispersion, constrained by the huge difference in the velocities between electrons and acoustic phonons, only optical phonons contribute to hot electron relaxation processes (see the schematic illustration in Figure~4d). However, for states around the Dirac nodal line without such restriction, acoustic phonons with momentum along the nodal line direction $q_\parallel$ can also take part in the relaxation process as the schematic illustration shown in Figure~4e, contributing additional scattering channels and thereby resulting in an even shorter relaxation time. We note while previous studies of a nodal-line semimetal ZrSiS have pointed out that acoustic phonon contributes to the relaxation process\cite{NeupaneZrSiS2022cp}, the role of the nodal-line electronic structure in the ultrafast relaxation dynamics has not been clarified yet.

To further demonstrate the role of acoustic phonons in the relaxation dynamics of the nodal lines in PtSn$_4$, we calculated the momentum-resolved el-ph coupling matrix for the electronic states around the Dirac nodal line.
As shown in Figure~4f, the cone-like electronic structure can be observed along $k_x$ with two linear bands, labeled as B1 and B2, crossing each other at the P point. Such a crossing corresponds to one gapless point on the nodal line (details in Figure~S3). Then, we calculate the intraband el-ph coupling matrix $\lambda_{q\nu k}^{intra}$ for the electronic states at the crossing P point and map the calculated results onto phonon bands as shown in Figure~4g,h. Consistent with discussions above, as final states near the nodal line, only optical phonons along the direction perpendicular to the nodal line ($q_\bot$) contribute to the el-ph scattering (Figure~4g and Figure~S4c), but when momentum is parallel to the nodal line ($q_{\parallel}$) both acoustic and optical phonon modes take part in the el-ph relaxations for the excited electron dynamics (Figure~4h and Figure~S4d).
Our calculated results confirm the analysis shown above and indicate the unique scattering mechanism in the Dirac nodal-line semimetal resulting in a faster relaxation. Such a conclusion is qualitatively consistent with the momentum and energy-resolved scattering time obtained from TrARPES measurements.

In summary, using TrARPES, we reveal the full evolution of electronic states around the curved Dirac nodal line near the $\overline{\rm Z}$ point which crosses the Fermi level in PtSn$_4$. By further monitoring the relaxation process of photoexcited electrons inside the projected nodal lines on the side surface, we find an ultrafast relaxation time as short as hundreds of femtoseconds, which results from the multichannel scatterings induced by electron-phonon coupling among the complex metallic bands of PtSn$_4$. Furthermore, a dynamical relaxation mechanism related to Dirac nodal-line semimetal is also identified, acting as an additional relaxation channel for photoexcited carriers. Our results provide a comprehensive understanding of the ultrafast dynamics in the Dirac nodal-line semimetals and reveal a unique scattering mechanism distinguished from other kinds of topological semimetals.


\section{METHODS}

\subsection{Sample Growth}

The single crystals of PtSn$_4$ were grown by the self-flux method. The starting elements, platinum powder and excess tin granules were put into the corundum crucible and sealed in a quartz tube with a ratio of Pt:Sn = 1:32. The quartz tube was heated to 600 $^{\circ}$C at 60 $^{\circ}$C/h and held for 24 h, then cooled to 380 $^{\circ}$C at a rate of 1 $^{\circ}$C/h, at which the excess Sn flux was separated from the crystals by centrifugation. The obtained crystals are thin plates with high quality.

\subsection{TrARPES Measurement and Analysis}

TrARPES measurements were performed in our home laboratory at Tsinghua University\cite{bch5d3_7RSI2022,ZHY100FSRSI2022}. The fundamental femtosecond laser pulses at 740 nm are provided by a Ti:sapphire oscillator together with a pulse picker running at a repetition rate of 4.75 MHz. The fundamental laser pulses are split into two beams, one used as the pump beam, and another used for generating the probe beam. The probe beam at 185 nm wavelength is obtained by two frequency doubling processes using nonlinear crystals of BBO and KBBF, respectively\cite{bch5d3_7RSI2022}. The pump fluence used in this paper is 150 $\mu$J/cm$^2$, and the time resolution is 150 $\pm$ 11 fs (see detail in Figure~S6). Single crystal PtSn$_4$ samples were cleaved \emph{in-situ} at 80 K with the base pressure better than $5 \times 10^{-11}$ Torr before TrARPES measurements.

The fitting function used to fit the relaxation time in Figure~3g-i is\cite{bch2021timeresolutionRSI, ZHY100FSRSI2022}: $$I(t) = A(1 + erf(2\sqrt{ln2}\frac{t-t_{0}}{\Delta t} - \frac{\Delta t}{4\sqrt{ln2}\tau}))e^{-\frac{t-t_{0}}{\tau}} + B.$$

\subsection{Calculation Method}

The electronic properties of bulk states without spin-orbital coupling (SOC) in PtSn$_4$ were calculated based on DFT calculation\cite{KohnPR1964,ShamPR1965} implemented in Vienna ab initio simulation package (VASP)\cite{FurthmullerPRB1996,FurthmullerCMS1996} (for the disccusion of SOC, see Figure~S7). The projector augmented wave (PAW) method\cite{BlochlPRB1994} and a plane-wave basis with a kinetic energy cutoff of 230 eV were employed to describe core-valence interactions. The Perdew-Burke-Ernzerhof (PBE) functional\cite{BlochlPRB1994} within the generalized gradient approximation was used to describe the exchange-correlation interactions. The BZ of the conventional cell was sampled by $8 \times 6 \times 8$ Monkhorst-Pack\cite{JamesPRB1976} $\bm{k}$-grids. The atomic coordinates were fully relaxed until the total energy difference was less than 1 meV. Lattice parameters of $a = 6.418$ {\AA}, $b = 11.366$ {\AA} and $c = 6.384$ {\AA} were used\cite{kaminskiARPES2016np}. The surface state calculations were carried out by using Wannier90\cite{wannierCPC2008} and WannierTool\cite{wannierCPC2018} packages. Maximally localized Wannier functions\cite{VanderbiltRMP2012} were used to describe the band structure around the Fermi level.

Calculations of dynamical matrices and phonon dispersions without SOC were performed based on density functional perturbation theory\cite{GiannozziRMP2001} implemented in {QUANTUM} {ESPRESSO} {(QE)} package\cite{WentzcovitchJPCM2009} accompanied by the PBE functional\cite{ErnzerhofPRL1996} and PAW method.\cite{DCCMS2014} The calculated results from QE and VASP packages are consistent. To reduce the computational costs, we used the primitive cell to calculate the el-ph matrix elements. The BZ of the primitive cell was sampled by $8 \times 8 \times 8$ coarse $\bm{k}$-grids and $2 \times 2 \times 2$ coarse $\bm{q}$-grids. The el-ph coupling strength was calculated on specific $\bm{k}$-points and $\bm{q}$-paths using the EPW code\cite{StevenPRB2007,GiustinoCPC2016} with the input from the QE package.

\begin{suppinfo}
	This material is available free of charge via the internet at http://pubs.acs.org. \\
	Dirac nodal lines around Z point, the comparison of the raw data and normalized data, the relaxation of excited electrons in PtSn$_4$, the time resolution of the system, influence of spin-orbital coupling effect (PDF).
\end{suppinfo}

\section{AUTHOR INFORMATION}
\subsection{Author Contributions}
$^{\infty}$ T.L. and Y.J. contributed equally to this work.

S.Z. initiated the research project. T.L. and H.Zhong performed the measurements and analyzed the data. X.Z., X.D., and T.X. grew the PtSn$_4$ single crystal. Y.J. and P.T. performed the theoretical analysis and calculation. T.L., Y.J., P.T., and S.Z. wrote the manuscript, and all authors contributed to the discussions and commented on the manuscript.

\subsection{Notes}
The authors declare no competing financial interest.

\begin{acknowledgement}
	This work is supported by the National Natural Science Foundation of China (Grant No.~12234011, 92250305, 12374053, 11725418, 11427903, 12074425), National Key R\&D Program of China (Grant No.~2021YFA1400100, 2019YFA0308602). C.B. is supported by a project funded by China Postdoctoral Science Foundation (No. BX20230187). C.B. and Hongyun Z. are supported by the Shuimu Tsinghua Scholar Program.
\end{acknowledgement}


\providecommand{\latin}[1]{#1}
\makeatletter
\providecommand{\doi}
{\begingroup\let\do\@makeother\dospecials
	\catcode`\{=1 \catcode`\}=2 \doi@aux}
\providecommand{\doi@aux}[1]{\endgroup\texttt{#1}}
\makeatother
\providecommand*\mcitethebibliography{\thebibliography}
\csname @ifundefined\endcsname{endmcitethebibliography}
{\let\endmcitethebibliography\endthebibliography}{}



\begin{mcitethebibliography}{52}
	\providecommand*\natexlab[1]{#1}
	\providecommand*\mciteSetBstSublistMode[1]{}
	\providecommand*\mciteSetBstMaxWidthForm[2]{}
	\providecommand*\mciteBstWouldAddEndPuncttrue
	{\def\EndOfBibitem{\unskip.}}
	\providecommand*\mciteBstWouldAddEndPunctfalse
	{\let\EndOfBibitem\relax}
	\providecommand*\mciteSetBstMidEndSepPunct[3]{}
	\providecommand*\mciteSetBstSublistLabelBeginEnd[3]{}
	\providecommand*\EndOfBibitem{}
	\mciteSetBstSublistMode{f}
	\mciteSetBstMaxWidthForm{subitem}{(\alph{mcitesubitemcount})}
	\mciteSetBstSublistLabelBeginEnd
	{\mcitemaxwidthsubitemform\space}
	{\relax}
	{\relax}
	
	\bibitem[Armitage \latin{et~al.}(2018)Armitage, Mele, and
	Vishwanath]{Armitagereview2018RMP}
	Armitage,~N.~P.; Mele,~E.~J.; Vishwanath,~A. {Weyl and Dirac semimetals in
		three-dimensional solids}. \emph{Rev. Mod. Phys.} \textbf{2018}, \emph{90},
	015001\relax
	\mciteBstWouldAddEndPuncttrue
	\mciteSetBstMidEndSepPunct{\mcitedefaultmidpunct}
	{\mcitedefaultendpunct}{\mcitedefaultseppunct}\relax
	\EndOfBibitem
	\bibitem[Lv \latin{et~al.}(2021)Lv, Qian, and Ding]{dinghongreviewRMP}
	Lv,~B.~Q.; Qian,~T.; Ding,~H. {Experimental perspective on three-dimensional
		topological semimetals}. \emph{Rev. Mod. Phys.} \textbf{2021}, \emph{93},
	025002\relax
	\mciteBstWouldAddEndPuncttrue
	\mciteSetBstMidEndSepPunct{\mcitedefaultmidpunct}
	{\mcitedefaultendpunct}{\mcitedefaultseppunct}\relax
	\EndOfBibitem
	\bibitem[Fang \latin{et~al.}(2016)Fang, Weng, Dai, and
	Fang]{FangzhongreviewCPB_2016}
	Fang,~C.; Weng,~H.; Dai,~X.; Fang,~Z. {Topological nodal line semimetals}.
	\emph{Chin. Phys. B} \textbf{2016}, \emph{25}, 117106\relax
	\mciteBstWouldAddEndPuncttrue
	\mciteSetBstMidEndSepPunct{\mcitedefaultmidpunct}
	{\mcitedefaultendpunct}{\mcitedefaultseppunct}\relax
	\EndOfBibitem
	\bibitem[Weber(2021)]{weberreview2021jap}
	Weber,~C.~P. {Ultrafast investigation and control of Dirac and Weyl
		semimetals}. \emph{J. Appl. Phys.} \textbf{2021}, \emph{129}, 070901\relax
	\mciteBstWouldAddEndPuncttrue
	\mciteSetBstMidEndSepPunct{\mcitedefaultmidpunct}
	{\mcitedefaultendpunct}{\mcitedefaultseppunct}\relax
	\EndOfBibitem
	\bibitem[Liu \latin{et~al.}(2020)Liu, Xia, Xiao, Garcia~de Abajo, and
	Sun]{sundong2020semimetals}
	Liu,~J.; Xia,~F.; Xiao,~D.; Garcia~de Abajo,~F.~J.; Sun,~D. {Semimetals for
		high-performance photodetection}. \emph{Nat. Mater.} \textbf{2020},
	\emph{19}, 830--837\relax
	\mciteBstWouldAddEndPuncttrue
	\mciteSetBstMidEndSepPunct{\mcitedefaultmidpunct}
	{\mcitedefaultendpunct}{\mcitedefaultseppunct}\relax
	\EndOfBibitem
	\bibitem[Ma \latin{et~al.}(2021)Ma, Grushin, and Burch]{ma2021topology}
	Ma,~Q.; Grushin,~A.~G.; Burch,~K.~S. {Topology and geometry under the nonlinear
		electromagnetic spotlight}. \emph{Nat. Mater.} \textbf{2021}, \emph{20},
	1601--1614\relax
	\mciteBstWouldAddEndPuncttrue
	\mciteSetBstMidEndSepPunct{\mcitedefaultmidpunct}
	{\mcitedefaultendpunct}{\mcitedefaultseppunct}\relax
	\EndOfBibitem
	\bibitem[Bao \latin{et~al.}(2022)Bao, Tang, Sun, and Zhou]{BCHreview2022NRP}
	Bao,~C.; Tang,~P.; Sun,~D.; Zhou,~S. {Light-induced emergent phenomena in 2D
		materials and topological materials}. \emph{Nat. Rev. Phys.} \textbf{2022},
	\emph{4}, 33--48\relax
	\mciteBstWouldAddEndPuncttrue
	\mciteSetBstMidEndSepPunct{\mcitedefaultmidpunct}
	{\mcitedefaultendpunct}{\mcitedefaultseppunct}\relax
	\EndOfBibitem
	\bibitem[Shin \latin{et~al.}(2024)Shin, Rubio, and Tang]{Dongbin2024}
	Shin,~D.; Rubio,~A.; Tang,~P. {Light-Induced Ideal Weyl Semimetal in HgTe via
		Nonlinear Phononics}. \emph{Phys. Rev. Lett.} \textbf{2024}, \emph{132},
	016603\relax
	\mciteBstWouldAddEndPuncttrue
	\mciteSetBstMidEndSepPunct{\mcitedefaultmidpunct}
	{\mcitedefaultendpunct}{\mcitedefaultseppunct}\relax
	\EndOfBibitem
	\bibitem[Ma \latin{et~al.}(2017)Ma, Xu, Chan, Zhang, Chang, Lin, Xie, Palacios,
	Lin, Jia, Lee, Jarillo-Herrero, and Gedik]{TaAsCPGE2017np}
	Ma,~Q.; Xu,~S.-Y.; Chan,~C.-K.; Zhang,~C.-L.; Chang,~G.; Lin,~Y.; Xie,~W.;
	Palacios,~T.; Lin,~H.; Jia,~S.; Lee,~P.~A.; Jarillo-Herrero,~P.; Gedik,~N.
	{Direct optical detection of Weyl fermion chirality in a topological
		semimetal}. \emph{Nat. Phys.} \textbf{2017}, \emph{13}, 842--847\relax
	\mciteBstWouldAddEndPuncttrue
	\mciteSetBstMidEndSepPunct{\mcitedefaultmidpunct}
	{\mcitedefaultendpunct}{\mcitedefaultseppunct}\relax
	\EndOfBibitem
	\bibitem[Osterhoudt \latin{et~al.}(2019)Osterhoudt, Diebel, Gray, Yang, Stanco,
	Huang, Shen, Ni, Moll, Ran, and S.]{TaAsshiftcurrent}
	Osterhoudt,~G.~B.; Diebel,~L.~K.; Gray,~M.~J.; Yang,~X.; Stanco,~J.; Huang,~X.;
	Shen,~B.; Ni,~N.; Moll,~P.~J.; Ran,~Y.; S.,~B.~K. {Colossal mid-infrared bulk
		photovoltaic effect in a type-I Weyl semimetal}. \emph{Nat. Mater.}
	\textbf{2019}, \emph{18}, 471--475\relax
	\mciteBstWouldAddEndPuncttrue
	\mciteSetBstMidEndSepPunct{\mcitedefaultmidpunct}
	{\mcitedefaultendpunct}{\mcitedefaultseppunct}\relax
	\EndOfBibitem
	\bibitem[Wu \latin{et~al.}(2017)Wu, Patankar, Morimoto, Nair, Thewalt, Little,
	Analytis, Moore, and Orenstein]{JosephSHG2017np}
	Wu,~L.; Patankar,~S.; Morimoto,~T.; Nair,~N.~L.; Thewalt,~E.; Little,~A.;
	Analytis,~J.~G.; Moore,~J.~E.; Orenstein,~J. {Giant anisotropic nonlinear
		optical response in transition metal monopnictide Weyl semimetals}.
	\emph{Nat. Phys.} \textbf{2017}, \emph{13}, 350--355\relax
	\mciteBstWouldAddEndPuncttrue
	\mciteSetBstMidEndSepPunct{\mcitedefaultmidpunct}
	{\mcitedefaultendpunct}{\mcitedefaultseppunct}\relax
	\EndOfBibitem
	\bibitem[Cheng \latin{et~al.}(2020)Cheng, Kanda, Ikeda, Matsuda, Xia, Schumann,
	Stemmer, Itatani, Armitage, and Matsunaga]{CdAsHHG2020PRL}
	Cheng,~B.; Kanda,~N.; Ikeda,~T.~N.; Matsuda,~T.; Xia,~P.; Schumann,~T.;
	Stemmer,~S.; Itatani,~J.; Armitage,~N.~P.; Matsunaga,~R. {Efficient Terahertz
		Harmonic Generation with Coherent Acceleration of Electrons in the Dirac
		Semimetal Cd$_3$As$_2$}. \emph{Phys. Rev. Lett.} \textbf{2020}, \emph{124},
	117402\relax
	\mciteBstWouldAddEndPuncttrue
	\mciteSetBstMidEndSepPunct{\mcitedefaultmidpunct}
	{\mcitedefaultendpunct}{\mcitedefaultseppunct}\relax
	\EndOfBibitem
	\bibitem[Bao \latin{et~al.}(2022)Bao, Li, Xu, Zhou, Zeng, Zhong, Gao, Luo, Sun,
	Xia, and Zhou]{bchpopulationinversion2022nl}
	Bao,~C.; Li,~Q.; Xu,~S.; Zhou,~S.; Zeng,~X.; Zhong,~H.; Gao,~Q.; Luo,~L.;
	Sun,~D.; Xia,~T.-L.; Zhou,~S. {Population Inversion and Dirac Fermion Cooling
		in 3D Dirac Semimetal Cd$_3$As$_2$}. \emph{Nano Lett.} \textbf{2022},
	\emph{22}, 1138--1144\relax
	\mciteBstWouldAddEndPuncttrue
	\mciteSetBstMidEndSepPunct{\mcitedefaultmidpunct}
	{\mcitedefaultendpunct}{\mcitedefaultseppunct}\relax
	\EndOfBibitem
	\bibitem[Liu \latin{et~al.}(2022)Liu, Dhakal, Sakhya, Beetar, Kabir, Regmi,
	Kaczorowski, Chini, Fregoso, and Neupane]{NeupaneZrSiS2022cp}
	Liu,~Y.; Dhakal,~G.; Sakhya,~A.~P.; Beetar,~J.~E.; Kabir,~F.; Regmi,~S.;
	Kaczorowski,~D.; Chini,~M.; Fregoso,~B.~M.; Neupane,~M. {Ultrafast relaxation
		of acoustic and optical phonons in a topological nodal-line semimetal ZrSiS}.
	\emph{Commun. Phys.} \textbf{2022}, \emph{5}, 203\relax
	\mciteBstWouldAddEndPuncttrue
	\mciteSetBstMidEndSepPunct{\mcitedefaultmidpunct}
	{\mcitedefaultendpunct}{\mcitedefaultseppunct}\relax
	\EndOfBibitem
	\bibitem[Weber \latin{et~al.}(2018)Weber, Schoop, Parkin, Newby, Nateprov,
	Lotsch, Mariserla, Kim, Dani, Bechtel, Arushanov, and
	Ali]{MazharZrSiS2018APL}
	Weber,~C.~P.; Schoop,~L.~M.; Parkin,~S. S.~P.; Newby,~R.~C.; Nateprov,~A.;
	Lotsch,~B.; Mariserla,~B. M.~K.; Kim,~J.~M.; Dani,~K.~M.; Bechtel,~H.~A.;
	Arushanov,~E.; Ali,~M. {Directly photoexcited Dirac and Weyl fermions in
		ZrSiS and NbAs}. \emph{Appl. Phys. Lett.} \textbf{2018}, \emph{113},
	221906\relax
	\mciteBstWouldAddEndPuncttrue
	\mciteSetBstMidEndSepPunct{\mcitedefaultmidpunct}
	{\mcitedefaultendpunct}{\mcitedefaultseppunct}\relax
	\EndOfBibitem
	\bibitem[Mun \latin{et~al.}(2012)Mun, Ko, Miller, Samolyuk, Bud'Ko, and
	Canfield]{firsttransport2012PRB}
	Mun,~E.; Ko,~H.; Miller,~G.~J.; Samolyuk,~G.~D.; Bud'Ko,~S.~L.; Canfield,~P.~C.
	{Magnetic field effects on transport properties of PtSn$_4$}. \emph{Phys.
		Rev. B} \textbf{2012}, \emph{85}, 035135\relax
	\mciteBstWouldAddEndPuncttrue
	\mciteSetBstMidEndSepPunct{\mcitedefaultmidpunct}
	{\mcitedefaultendpunct}{\mcitedefaultseppunct}\relax
	\EndOfBibitem
	\bibitem[Yan \latin{et~al.}(2020)Yan, Luo, Gao, Lv, Xi, Sun, Lu, Tong, Sheng,
	Zhu, Song, and Y.P.]{PHEYan2020JPCM}
	Yan,~J.; Luo,~X.; Gao,~J.; Lv,~H.; Xi,~C.; Sun,~Y.; Lu,~W.; Tong,~P.;
	Sheng,~Z.; Zhu,~X.; Song,~W.; Y.P.,~S. {The giant planar Hall effect and
		anisotropic magnetoresistance in Dirac node arcs semimetal PtSn$_4$}.
	\emph{J. Phys.: Condens. Matter} \textbf{2020}, \emph{32}, 315702\relax
	\mciteBstWouldAddEndPuncttrue
	\mciteSetBstMidEndSepPunct{\mcitedefaultmidpunct}
	{\mcitedefaultendpunct}{\mcitedefaultseppunct}\relax
	\EndOfBibitem
	\bibitem[Yara \latin{et~al.}(2018)Yara, Kakihana, Nishimura, Hedo, Nakama,
	{\=O}nuki, and Harima]{HarimaPS2018PBCM}
	Yara,~T.; Kakihana,~M.; Nishimura,~K.; Hedo,~M.; Nakama,~T.; {\=O}nuki,~Y.;
	Harima,~H. {Small fermi surfaces of PtSn$_4$ and Pt$_3$In$_7$}. \emph{Phys.
		B} \textbf{2018}, \emph{536}, 625--633\relax
	\mciteBstWouldAddEndPuncttrue
	\mciteSetBstMidEndSepPunct{\mcitedefaultmidpunct}
	{\mcitedefaultendpunct}{\mcitedefaultseppunct}\relax
	\EndOfBibitem
	\bibitem[Wang \latin{et~al.}(2018)Wang, Liang, Ge, Yang, Gong, Luo, Pi, Zhu,
	Zhang, and Zhang]{ZhangYHPSdhva2018JPCM}
	Wang,~Y.~J.; Liang,~D.~D.; Ge,~M.; Yang,~J.; Gong,~J.~X.; Luo,~L.; Pi,~L.;
	Zhu,~W.~K.; Zhang,~C.~J.; Zhang,~Y.~H. {Topological nature of the node-arc
		semimetal PtSn$_4$ probed by de Haas-van Alphen quantum oscillations}.
	\emph{J. Phys.: Condens. Matter} \textbf{2018}, \emph{30}, 155701\relax
	\mciteBstWouldAddEndPuncttrue
	\mciteSetBstMidEndSepPunct{\mcitedefaultmidpunct}
	{\mcitedefaultendpunct}{\mcitedefaultseppunct}\relax
	\EndOfBibitem
	\bibitem[Luo \latin{et~al.}(2018)Luo, Xiao, Chen, Yan, Pei, Sun, Lu, Tong,
	Sheng, Zhu, Song, and Sun]{SunYPPSLifshtz2018PRB}
	Luo,~X.; Xiao,~R.~C.; Chen,~F.~C.; Yan,~J.; Pei,~Q.~L.; Sun,~Y.; Lu,~W.~J.;
	Tong,~P.; Sheng,~Z.~G.; Zhu,~X.~B.; Song,~W.~H.; Sun,~Y.~P. {Origin of the
		extremely large magnetoresistance in topological semimetal PtSn$_4$}.
	\emph{Phys. Rev. B} \textbf{2018}, \emph{97}, 205132\relax
	\mciteBstWouldAddEndPuncttrue
	\mciteSetBstMidEndSepPunct{\mcitedefaultmidpunct}
	{\mcitedefaultendpunct}{\mcitedefaultseppunct}\relax
	\EndOfBibitem
	\bibitem[Marchenkov \latin{et~al.}(2019)Marchenkov, Domozhirova, Makhnev,
	Shreder, Lukoyanov, Naumov, Chistyakov, Marchenkova, Huang, and
	Eisterer]{Eisterer2019JOEATP}
	Marchenkov,~V.~V.; Domozhirova,~A.~N.; Makhnev,~A.~A.; Shreder,~E.~I.;
	Lukoyanov,~A.~V.; Naumov,~S.~V.; Chistyakov,~V.~V.; Marchenkova,~E.~B.;
	Huang,~J. C.~A.; Eisterer,~M. {Electronic Structure and Electronic Properties
		of PtSn$_4$ Single Crystal}. \emph{J. Exp. Theor. Phys.} \textbf{2019},
	\emph{128}, 939--945\relax
	\mciteBstWouldAddEndPuncttrue
	\mciteSetBstMidEndSepPunct{\mcitedefaultmidpunct}
	{\mcitedefaultendpunct}{\mcitedefaultseppunct}\relax
	\EndOfBibitem
	\bibitem[Fu \latin{et~al.}(2020)Fu, Guin, Scaffidi, Sun, Saha, Watzman,
	Srivastava, Li, Schnelle, Parkin, Felser, and Gooth]{fuchenguang2020Research}
	Fu,~C.; Guin,~S.~N.; Scaffidi,~T.; Sun,~Y.; Saha,~R.; Watzman,~S.~J.;
	Srivastava,~A.~K.; Li,~G.; Schnelle,~W.; Parkin,~S.~S.; Felser,~C.; Gooth,~J.
	{Largely suppressed Magneto-Thermal Conductivity and Enhanced
		Magneto-Thermoelectric Properties in PtSn$_4$}. \emph{Research}
	\textbf{2020}, \emph{2020}, 4643507\relax
	\mciteBstWouldAddEndPuncttrue
	\mciteSetBstMidEndSepPunct{\mcitedefaultmidpunct}
	{\mcitedefaultendpunct}{\mcitedefaultseppunct}\relax
	\EndOfBibitem
	\bibitem[Wu \latin{et~al.}(2016)Wu, Wang, Mun, Johnson, Mou, Huang, Lee,
	Bud’ko, Canfield, and Kaminski]{kaminskiARPES2016np}
	Wu,~Y.; Wang,~L.-L.; Mun,~E.; Johnson,~D.~D.; Mou,~D.; Huang,~L.; Lee,~Y.;
	Bud’ko,~S.~L.; Canfield,~P.~C.; Kaminski,~A. {Dirac node arcs in PtSn$_4$}.
	\emph{Nat. Phys.} \textbf{2016}, \emph{12}, 667--671\relax
	\mciteBstWouldAddEndPuncttrue
	\mciteSetBstMidEndSepPunct{\mcitedefaultmidpunct}
	{\mcitedefaultendpunct}{\mcitedefaultseppunct}\relax
	\EndOfBibitem
	\bibitem[Bao \latin{et~al.}(2021)Bao, Luo, Zhang, Zhou, Ren, and
	Zhou]{bch2021timeresolutionRSI}
	Bao,~C.; Luo,~L.; Zhang,~H.; Zhou,~S.; Ren,~Z.; Zhou,~S. {Full diagnostics and
		optimization of time resolution for time-and angle-resolved photoemission
		spectroscopy}. \emph{Rev. Sci. Instrum.} \textbf{2021}, \emph{92},
	033904\relax
	\mciteBstWouldAddEndPuncttrue
	\mciteSetBstMidEndSepPunct{\mcitedefaultmidpunct}
	{\mcitedefaultendpunct}{\mcitedefaultseppunct}\relax
	\EndOfBibitem
	\bibitem[Zhong \latin{et~al.}(2022)Zhong, Bao, Lin, Zhou, and
	Zhou]{ZHY100FSRSI2022}
	Zhong,~H.; Bao,~C.; Lin,~T.; Zhou,~S.; Zhou,~S. {A newly designed femtosecond
		KBe$_2$BO$_3$F$_2$ device with pulse duration down to 55 fs for time- and
		angle-resolved photoemission spectroscopy}. \emph{Rev. Sci. Instrum.}
	\textbf{2022}, \emph{93}, 113910\relax
	\mciteBstWouldAddEndPuncttrue
	\mciteSetBstMidEndSepPunct{\mcitedefaultmidpunct}
	{\mcitedefaultendpunct}{\mcitedefaultseppunct}\relax
	\EndOfBibitem
	\bibitem[Sobota \latin{et~al.}(2012)Sobota, Yang, Analytis, Chen, Fisher,
	Kirchmann, and Shen]{shenZXBS2012PRL}
	Sobota,~J.~A.; Yang,~S.; Analytis,~J.~G.; Chen,~Y.; Fisher,~I.~R.;
	Kirchmann,~P.~S.; Shen,~Z.-X. {Ultrafast Optical Excitation of a Persistent
		Surface-State Population in the Topological Insulator Bi$_2$Se$_3$}.
	\emph{Phys. Rev. Lett.} \textbf{2012}, \emph{108}, 117403\relax
	\mciteBstWouldAddEndPuncttrue
	\mciteSetBstMidEndSepPunct{\mcitedefaultmidpunct}
	{\mcitedefaultendpunct}{\mcitedefaultseppunct}\relax
	\EndOfBibitem
	\bibitem[Wang \latin{et~al.}(2012)Wang, Hsieh, Sie, Steinberg, Gardner, Lee,
	Jarillo-Herrero, and Gedik]{GedikBS2012PRL}
	Wang,~Y.; Hsieh,~D.; Sie,~E.; Steinberg,~H.; Gardner,~D.; Lee,~Y.;
	Jarillo-Herrero,~P.; Gedik,~N. {Measurement of Intrinsic Dirac Fermion
		Cooling on the Surface of the Topological Insulator Bi$_2$Se$_3$ Using
		Time-Resolved and Angle-Resolved Photoemission Spectroscopy}. \emph{Phys.
		Rev. Lett.} \textbf{2012}, \emph{109}, 127401\relax
	\mciteBstWouldAddEndPuncttrue
	\mciteSetBstMidEndSepPunct{\mcitedefaultmidpunct}
	{\mcitedefaultendpunct}{\mcitedefaultseppunct}\relax
	\EndOfBibitem
	\bibitem[Reimann \latin{et~al.}(2014)Reimann, G\"udde, Kuroda, Chulkov, and
	H\"ofer]{HoferSbTe2014PRB}
	Reimann,~J.; G\"udde,~J.; Kuroda,~K.; Chulkov,~E.~V.; H\"ofer,~U. {Spectroscopy
		and dynamics of unoccupied electronic states of the topological insulators
		Sb$_2$Te$_{3}$ and Sb$_{2}$Te$_{2}$S}. \emph{Phys. Rev. B} \textbf{2014},
	\emph{90}, 081106\relax
	\mciteBstWouldAddEndPuncttrue
	\mciteSetBstMidEndSepPunct{\mcitedefaultmidpunct}
	{\mcitedefaultendpunct}{\mcitedefaultseppunct}\relax
	\EndOfBibitem
	\bibitem[Zhu \latin{et~al.}(2015)Zhu, Ishida, Kuroda, Sumida, Ye, Wang, Pan,
	Taniguchi, Qiao, Shin, and Kimura]{KimuraSbTe2015scirep}
	Zhu,~S.; Ishida,~Y.; Kuroda,~K.; Sumida,~K.; Ye,~M.; Wang,~J.; Pan,~H.;
	Taniguchi,~M.; Qiao,~S.; Shin,~S.; Kimura,~A. {Ultrafast electron dynamics at
		the Dirac node of the topological insulator Sb$_2$Te$_3$}. \emph{Sci. Rep.}
	\textbf{2015}, \emph{5}, 13213\relax
	\mciteBstWouldAddEndPuncttrue
	\mciteSetBstMidEndSepPunct{\mcitedefaultmidpunct}
	{\mcitedefaultendpunct}{\mcitedefaultseppunct}\relax
	\EndOfBibitem
	\bibitem[Zhong \latin{et~al.}(2021)Zhong, Bao, Wang, Li, Yin, Xu, Duan, Xia,
	and Zhou]{zhyMBT2021nl}
	Zhong,~H.; Bao,~C.; Wang,~H.; Li,~J.; Yin,~Z.; Xu,~Y.; Duan,~W.; Xia,~T.-L.;
	Zhou,~S. {Light-Tunable Surface State and Hybridization Gap in Magnetic
		Topological Insulator MnBi$_8$Te$_{13}$}. \emph{Nano Lett.} \textbf{2021},
	\emph{21}, 6080--6086\relax
	\mciteBstWouldAddEndPuncttrue
	\mciteSetBstMidEndSepPunct{\mcitedefaultmidpunct}
	{\mcitedefaultendpunct}{\mcitedefaultseppunct}\relax
	\EndOfBibitem
	\bibitem[Majchrzak \latin{et~al.}(2023)Majchrzak, Liu, Volckaert, Biswas,
	Sahoo, Puntel, Bronsch, Tuniz, Cilento, Pan, Liu, Chen, and
	Soren]{majchrzakMBT2023nl}
	Majchrzak,~P.~E.; Liu,~Y.; Volckaert,~K.; Biswas,~D.; Sahoo,~C.; Puntel,~D.;
	Bronsch,~W.; Tuniz,~M.; Cilento,~F.; Pan,~X.-C.; Liu,~Q.; Chen,~Y.~P.;
	Soren,~U. {Van der Waals Engineering of Ultrafast Carrier Dynamics in
		Magnetic Heterostructures}. \emph{Nano Lett.} \textbf{2023}, \emph{23},
	414--421\relax
	\mciteBstWouldAddEndPuncttrue
	\mciteSetBstMidEndSepPunct{\mcitedefaultmidpunct}
	{\mcitedefaultendpunct}{\mcitedefaultseppunct}\relax
	\EndOfBibitem
	\bibitem[Xue \latin{et~al.}(2022)Xue, Liao, Li, Hu, Chen, Tang, Yang, Liao, Lu,
	and Gong]{gongqihuangCdAs2022JPCC}
	Xue,~Z.; Liao,~X.; Li,~Y.; Hu,~A.; Chen,~J.; Tang,~J.; Yang,~H.; Liao,~Z.-M.;
	Lu,~G.; Gong,~Q. {Photoexcited Electron Dynamics in Cd$_3$As$_2$ Revealed by
		Time-and Energy-Resolved Photoemission Electron Microscopy}. \emph{J. Phys.
		Chem. C} \textbf{2022}, \emph{126}, 3134--3139\relax
	\mciteBstWouldAddEndPuncttrue
	\mciteSetBstMidEndSepPunct{\mcitedefaultmidpunct}
	{\mcitedefaultendpunct}{\mcitedefaultseppunct}\relax
	\EndOfBibitem
	\bibitem[Gierz \latin{et~al.}(2013)Gierz, Petersen, Mitrano, Cacho, Turcu,
	Springate, St{\"o}hr, K{\"o}hler, Starke, and
	Cavalleri]{Andreagraphene2013nm}
	Gierz,~I.; Petersen,~J.~C.; Mitrano,~M.; Cacho,~C.; Turcu,~I.~E.;
	Springate,~E.; St{\"o}hr,~A.; K{\"o}hler,~A.; Starke,~U.; Cavalleri,~A.
	{Snapshots of non-equilibrium Dirac carrier distributions in graphene}.
	\emph{Nat. Mater.} \textbf{2013}, \emph{12}, 1119--1124\relax
	\mciteBstWouldAddEndPuncttrue
	\mciteSetBstMidEndSepPunct{\mcitedefaultmidpunct}
	{\mcitedefaultendpunct}{\mcitedefaultseppunct}\relax
	\EndOfBibitem
	\bibitem[Gierz \latin{et~al.}(2015)Gierz, Mitrano, Petersen, Cacho, Turcu,
	Springate, Stöhr, Köhler, Starke, and Cavalleri]{Andreagraphene2015}
	Gierz,~I.; Mitrano,~M.; Petersen,~J.~C.; Cacho,~C.; Turcu,~I. C.~E.;
	Springate,~E.; Stöhr,~A.; Köhler,~A.; Starke,~U.; Cavalleri,~A. {Population
		inversion in monolayer and bilayer graphene}. \emph{J. Phys.: Condens.
		Matter} \textbf{2015}, \emph{27}, 164204\relax
	\mciteBstWouldAddEndPuncttrue
	\mciteSetBstMidEndSepPunct{\mcitedefaultmidpunct}
	{\mcitedefaultendpunct}{\mcitedefaultseppunct}\relax
	\EndOfBibitem
	\bibitem[Brida \latin{et~al.}(2013)Brida, Tomadin, Manzoni, Kim, Lombardo,
	Milana, Nair, Novoselov, Ferrari, Cerullo, and Polini]{Polinigraphene2013nc}
	Brida,~D.; Tomadin,~A.; Manzoni,~C.; Kim,~Y.; Lombardo,~A.; Milana,~S.;
	Nair,~R.; Novoselov,~K.; Ferrari,~A.; Cerullo,~G.; Polini,~M. Ultrafast
	collinear scattering and carrier multiplication in graphene. \emph{Nat.
		Commun.} \textbf{2013}, \emph{4}, 1987\relax
	\mciteBstWouldAddEndPuncttrue
	\mciteSetBstMidEndSepPunct{\mcitedefaultmidpunct}
	{\mcitedefaultendpunct}{\mcitedefaultseppunct}\relax
	\EndOfBibitem
	\bibitem[Bao \latin{et~al.}(2022)Bao, Zhong, Zhou, Feng, Wang, and
	Zhou]{bch5d3_7RSI2022}
	Bao,~C.; Zhong,~H.; Zhou,~S.; Feng,~R.; Wang,~Y.; Zhou,~S. {Ultrafast time- and
		angle-resolved photoemission spectroscopy with widely tunable probe photon
		energy of 5.3--7.0 eV for investigating dynamics of three-dimensional
		materials}. \emph{Rev. Sci. Instrum.} \textbf{2022}, \emph{93}, 013902\relax
	\mciteBstWouldAddEndPuncttrue
	\mciteSetBstMidEndSepPunct{\mcitedefaultmidpunct}
	{\mcitedefaultendpunct}{\mcitedefaultseppunct}\relax
	\EndOfBibitem
	\bibitem[Hohenberg and Kohn(1964)Hohenberg, and Kohn]{KohnPR1964}
	Hohenberg,~P.; Kohn,~W. {Inhomogeneous Electron Gas}. \emph{Phys. Rev.}
	\textbf{1964}, \emph{136}, B864--B871\relax
	\mciteBstWouldAddEndPuncttrue
	\mciteSetBstMidEndSepPunct{\mcitedefaultmidpunct}
	{\mcitedefaultendpunct}{\mcitedefaultseppunct}\relax
	\EndOfBibitem
	\bibitem[Kohn and Sham(1965)Kohn, and Sham]{ShamPR1965}
	Kohn,~W.; Sham,~L.~J. {Self-Consistent Equations Including Exchange and
		Correlation Effects}. \emph{Phys. Rev.} \textbf{1965}, \emph{140},
	A1133--A1138\relax
	\mciteBstWouldAddEndPuncttrue
	\mciteSetBstMidEndSepPunct{\mcitedefaultmidpunct}
	{\mcitedefaultendpunct}{\mcitedefaultseppunct}\relax
	\EndOfBibitem
	\bibitem[Kresse and Furthmüller(1996)Kresse, and
	Furthmüller]{FurthmullerPRB1996}
	Kresse,~G.; Furthmüller,~J. {Efficient iterative schemes for ab initio
		total-energy calculations using a plane-wave basis set}. \emph{Phys. Rev. B}
	\textbf{1996}, \emph{54}, 11169--11186\relax
	\mciteBstWouldAddEndPuncttrue
	\mciteSetBstMidEndSepPunct{\mcitedefaultmidpunct}
	{\mcitedefaultendpunct}{\mcitedefaultseppunct}\relax
	\EndOfBibitem
	\bibitem[Kresse and Furthmüller(1996)Kresse, and
	Furthmüller]{FurthmullerCMS1996}
	Kresse,~G.; Furthmüller,~J. {Efficiency of ab-initio total energy calculations
		for metals and semiconductors using a plane-wave basis set}. \emph{Comput.
		Mater. Sci.} \textbf{1996}, \emph{6}, 15--50\relax
	\mciteBstWouldAddEndPuncttrue
	\mciteSetBstMidEndSepPunct{\mcitedefaultmidpunct}
	{\mcitedefaultendpunct}{\mcitedefaultseppunct}\relax
	\EndOfBibitem
	\bibitem[Blöchl(1994)]{BlochlPRB1994}
	Blöchl,~P.~E. {Projector augmented-wave method}. \emph{Phys. Rev. B}
	\textbf{1994}, \emph{50}, 17953--17979\relax
	\mciteBstWouldAddEndPuncttrue
	\mciteSetBstMidEndSepPunct{\mcitedefaultmidpunct}
	{\mcitedefaultendpunct}{\mcitedefaultseppunct}\relax
	\EndOfBibitem
	\bibitem[Monkhorst and Pack(1976)Monkhorst, and Pack]{JamesPRB1976}
	Monkhorst,~H.~J.; Pack,~J.~D. {Special points for Brillouin-zone integrations}.
	\emph{Phys. Rev. B} \textbf{1976}, \emph{13}, 5188--5192\relax
	\mciteBstWouldAddEndPuncttrue
	\mciteSetBstMidEndSepPunct{\mcitedefaultmidpunct}
	{\mcitedefaultendpunct}{\mcitedefaultseppunct}\relax
	\EndOfBibitem
	\bibitem[Mostofi \latin{et~al.}(2008)Mostofi, Yates, Lee, Souza, Vanderbilt,
	and Marzari]{wannierCPC2008}
	Mostofi,~A.~A.; Yates,~J.~R.; Lee,~Y.-S.; Souza,~I.; Vanderbilt,~D.;
	Marzari,~N. {wannier90: A tool for obtaining maximally-localised Wannier
		functions}. \emph{Comput. Phys. Commun.} \textbf{2008}, \emph{178},
	685--699\relax
	\mciteBstWouldAddEndPuncttrue
	\mciteSetBstMidEndSepPunct{\mcitedefaultmidpunct}
	{\mcitedefaultendpunct}{\mcitedefaultseppunct}\relax
	\EndOfBibitem
	\bibitem[Wu \latin{et~al.}(2018)Wu, Zhang, Song, Troyer, and
	Soluyanov]{wannierCPC2018}
	Wu,~Q.; Zhang,~S.; Song,~H.-F.; Troyer,~M.; Soluyanov,~A.~A. {WannierTools: An
		open-source software package for novel topological materials}. \emph{Comput.
		Phys. Commun.} \textbf{2018}, \emph{224}, 405--416\relax
	\mciteBstWouldAddEndPuncttrue
	\mciteSetBstMidEndSepPunct{\mcitedefaultmidpunct}
	{\mcitedefaultendpunct}{\mcitedefaultseppunct}\relax
	\EndOfBibitem
	\bibitem[Marzari \latin{et~al.}(2012)Marzari, Mostofi, Yates, Souza, and
	Vanderbilt]{VanderbiltRMP2012}
	Marzari,~N.; Mostofi,~A.~A.; Yates,~J.~R.; Souza,~I.; Vanderbilt,~D. {Maximally
		localized Wannier functions: Theory and applications}. \emph{Rev. Mod. Phys.}
	\textbf{2012}, \emph{84}, 1419--1475\relax
	\mciteBstWouldAddEndPuncttrue
	\mciteSetBstMidEndSepPunct{\mcitedefaultmidpunct}
	{\mcitedefaultendpunct}{\mcitedefaultseppunct}\relax
	\EndOfBibitem
	\bibitem[Baroni \latin{et~al.}(2001)Baroni, de~Gironcoli, Dal~Corso, and
	Giannozzi]{GiannozziRMP2001}
	Baroni,~S.; de~Gironcoli,~S.; Dal~Corso,~A.; Giannozzi,~P. {Phonons and related
		crystal properties from density-functional perturbation theory}. \emph{Rev.
		Mod. Phys.} \textbf{2001}, \emph{73}, 515--562\relax
	\mciteBstWouldAddEndPuncttrue
	\mciteSetBstMidEndSepPunct{\mcitedefaultmidpunct}
	{\mcitedefaultendpunct}{\mcitedefaultseppunct}\relax
	\EndOfBibitem
	\bibitem[Giannozzi \latin{et~al.}(2009)Giannozzi, Baroni, Bonini, Calandra,
	Car, Cavazzoni, Ceresoli, Chiarotti, Cococcioni, Dabo, Dal~Corso,
	de~Gironcoli, Fabris, Fratesi, Gebauer, Gerstmann, Gougoussis, Kokalj,
	Lazzeri, Martin-Samos, Marzari, Mauri, Mazzarello, Paolini, Pasquarello,
	Paulatto, Sbraccia, Scandolo, Sclauzero, Seitsonen, Smogunov, Umari, and
	Wentzcovitch]{WentzcovitchJPCM2009}
	Giannozzi,~P. \latin{et~al.}  {QUANTUM ESPRESSO: a modular and open-source
		software project for quantum simulations of materials}. \emph{J. Phys.:
		Condens. Matter} \textbf{2009}, \emph{21}, 395502\relax
	\mciteBstWouldAddEndPuncttrue
	\mciteSetBstMidEndSepPunct{\mcitedefaultmidpunct}
	{\mcitedefaultendpunct}{\mcitedefaultseppunct}\relax
	\EndOfBibitem
	\bibitem[Perdew \latin{et~al.}(1996)Perdew, Burke, and
	Ernzerhof]{ErnzerhofPRL1996}
	Perdew,~J.~P.; Burke,~K.; Ernzerhof,~M. {Generalized Gradient Approximation
		Made Simple}. \emph{Phys. Rev. Lett.} \textbf{1996}, \emph{77},
	3865--3868\relax
	\mciteBstWouldAddEndPuncttrue
	\mciteSetBstMidEndSepPunct{\mcitedefaultmidpunct}
	{\mcitedefaultendpunct}{\mcitedefaultseppunct}\relax
	\EndOfBibitem
	\bibitem[Dal~Corso(2014)]{DCCMS2014}
	Dal~Corso,~A. {Pseudopotentials periodic table: From H to Pu}. \emph{Comput.
		Mater. Sci.} \textbf{2014}, \emph{95}, 337--350\relax
	\mciteBstWouldAddEndPuncttrue
	\mciteSetBstMidEndSepPunct{\mcitedefaultmidpunct}
	{\mcitedefaultendpunct}{\mcitedefaultseppunct}\relax
	\EndOfBibitem
	\bibitem[Giustino \latin{et~al.}(2007)Giustino, Cohen, and
	Louie]{StevenPRB2007}
	Giustino,~F.; Cohen,~M.~L.; Louie,~S.~G. {Electron-phonon interaction using
		Wannier functions}. \emph{Phys. Rev. B} \textbf{2007}, \emph{76},
	165108\relax
	\mciteBstWouldAddEndPuncttrue
	\mciteSetBstMidEndSepPunct{\mcitedefaultmidpunct}
	{\mcitedefaultendpunct}{\mcitedefaultseppunct}\relax
	\EndOfBibitem
	\bibitem[Poncé \latin{et~al.}(2016)Poncé, Margine, Verdi, and
	Giustino]{GiustinoCPC2016}
	Poncé,~S.; Margine,~E.~R.; Verdi,~C.; Giustino,~F. {EPW: Electron–phonon
		coupling, transport and superconducting properties using maximally localized
		Wannier functions}. \emph{Comput. Phys. Commun.} \textbf{2016}, \emph{209},
	116--133\relax
	\mciteBstWouldAddEndPuncttrue
	\mciteSetBstMidEndSepPunct{\mcitedefaultmidpunct}
	{\mcitedefaultendpunct}{\mcitedefaultseppunct}\relax
	\EndOfBibitem
\end{mcitethebibliography}
\end{document}